# Ambivalence in stakeholders' views on connected and autonomous vehicles

Celina Kacperski[1]   Tobias Vogel[1]   Florian Kutzner[1]

[1] Mannheim University, Germany

**Abstract.** Connected and autonomous vehicles (CAVs) are often discussed as a solution to pressing issues of the current transport systems, including congestion, safety, social inclusion and ecological sustainability. Scientifically, there is agreement that CAVs may solve, but can also aggravate these issues, depending on the specific CAV solution. In the current paper, we investigate the visions and worst-case scenarios of various stakeholders, including representatives of public administrations, automotive original equipment manufacturers, insurance companies, public transportation service providers, mobility experts and politicians. A qualitative analysis of 17 semi-structured interviews is presented. It reveals experts' ambivalence towards the introduction of CAVs, reflecting high levels of uncertainty about CAV consequences, including issues of efficiency, comfort and sustainability, and concerns about co-road users such as pedestrians and cyclists. Implications of the sluggishness of policymakers to set boundary conditions and for the labor market are discussed. An open debate between policymakers, citizens and other stakeholders on how to introduce CAVs seems timely.

**Note:** This research was funded by the H2020 PAsCAL project (grant agreement number 815098). We would like to thank Luca Pier for her help with interview transcription and coding.

## 1    Introduction

Mobility and transportation systems, as they currently operate, are socially and environmentally unsustainable (Burns, 2013). The development of advanced vehicle technologies and alternative fuel types has the potential to positively affect both humans and the environment by enhancing driving experience, making it more socially inclusive, and reducing the carbon footprint of the transport system (Greenblatt & Shaheen, 2015; Kirk & Eng, 2011; Litman, 2019). The evolution of connected and autonomous vehicles (CAVs) is one central part of this development. Yet, while the number of testbeds and exemptions for on-street use of fully autonomous vehicles are increasing (Innamaa, 2019; Lee, 2020), there does not seem to be a coherent vision as to how CAVs are going to be integrated into the mobility eco-system.

Yet, the consequences of the integration of CAVs need to be carefully considered. The current literature on CAVs has started to outline positive and negative consequences of large-scale CAV adoption (for a recent review see, Narayanan et al., 2020). In terms of traffic and travel behavior, less disutility of travel time could lead to rebounds, increasing road network load potentially offsetting the initial benefits (Medina-Tapia & Robusté, 2019; Taiebat et al., 2019). In terms of safety, estimates predict an accident reduction by a third if all vehicles had forward collision and lane departure warning systems, sideview (blind spot) assist, and adaptive headlights, with the main reduction being introduced by Level 4 automation onwards (IIHS, 2010). However, due to lack of real-world data, studies have mostly used simulations to arrive at these numbers (Papadoulis et al., 2019). The ample potential for malevolent outside influences to severely reduce safety has been discussed (Parkinson et al., 2017).

In terms of environmental consequences, multitudes of aspects are of note. CAVs have been predicted to improve fuel economy per kilometer travelled through smoother acceleration and tighter platooning, with higher effective speeds (Anderson et al., 2016). Integrated into taxi or sharing operations, the average number of people per vehicle could increase and the average size of the vehicle, and possibly its battery, decrease when adapted to the real occupancy (Burns, 2013; Burns et al., 2013; Shiau et al., 2009). At the same time, increased travel demand, increased infrastructure need for communications, vehicle to vehicle (V2V) and vehicle to infrastructure (V2I), the inclusion of new user groups, and a cannibalizing effect on public transport might limit - or even reverse - these positive environmental impacts (Anderson et al., 2016; Greenblatt & Shaheen, 2015; Taiebat et al., 2018, 2019; Wadud et al., 2016). Expected impacts on land use with the associated loss in biodiversity are similarly heterogeneous. While the need for parking space might be severely reduced (Zhang & Guhathakurta, 2017), urban sprawl might further intensify especially at higher levels of automation (Zhang & Guhathakurta, 2018).

The potential social impacts of the integration of CAVs are at least as manifold as the environmental ones. CAVs offer obvious benefits to the blind and partially sighted, to the elderly and underaged, and to the physically or mentally challenged (Harrison & Ragland, 2003; Taylor & Tripodes, 2001). Relying on CAVs, these groups could enjoy unprecedented freedom of movement. Yet, social inclusion hinges on several factors, such as user interfaces and vehicles being designed to meet the diverse needs and the availability of CAVs at a reasonable cost; both seem somewhat questionable given the current focus on traditional business models (Arieff, 2013). Further, economic disruptions, including job creation and losses, can be expected for parts of the industry. Car manufacturing will undergo changes that are hard to predict, while driving and "crash economy" related jobs will be lost (Anderson et al., 2016).

Given the heterogeneity of possible consequences, the simplicity with which previous research has looked at the acceptance of CAVs is noteworthy. Most surveys have left the type of CAV and its usage unspecified, or supplied minimal information about level of automation and ownership (Bansal et al., 2016; Haboucha et al., 2017; Kyriakidis et al., 2015; Schoettle & Sivak, 2014) or focused on a single specific solution, such as for example a small autonomous shuttle bus (Nordhoff et al., 2018). Additionally, no study has provided information on the diverse possible consequences in relation to acceptance. Potential future consumers might have had very little information on which to base survey or interview responses. The simple dimensional structure of CAV acceptance thus might reflect a general attitude towards novel technologies in combination with specific concerns about security and legal issues (Nordhoff et al., 2018;

Payre et al., 2014). After all, assessing acceptance towards a single CAV solution is unlikely to provide a picture suitable to illustrate the diverse facets of CAV acceptance.

However, a comprehensive description of CAVs, capturing the variety of both, so-lutions and consequences, seems to be an unrealistically overambitious endeavour. To account for the variety on the one hand, and the uncertainty on the other, we therefore seek to approach CAV acceptance in a qualitative manner. For this purpose, we investigate vision and worst-case scenarios held by representatives of stakeholder groups that will be shaping how CAVs are introduced. So far, stakeholder evaluations and their acceptance of CAVs have rarely been sought systematically: when expert stakeholders were the target of research, it was to source the time horizon for the introduction of various levels of automation on the roads (Underwood & Firmin, 2014). The current paper seeks to inform research on the introduction of CAVs by sourcing the knowledge and visions of stakeholders in the field. We expect to find large variability in experts' views and, in an exploratory fashion, will investigate whether there are shared vision and worst-case scenarios, and barriers to adoption in relation to those.

## 2    Methods

### 2.1    Participants

Recruitment was carried out via email outreach to those representatives with specific stakeholder expertise (as listed in Table 1, Stakeholder Group).

**Table 1.** Stakeholders and their expertise.

| Pseudonym | Stakeholder Group | Main Expertise |
|-----------|-------------------|----------------|
| A1 – A4 | Academics | Mobility simulations; home-driving simulators; autonomous vehicle acceptance |
| C1 – C6 | Mobility consultants and associations | Public transport; driving school; peer-to-peer mobility and crowdsourced mobility |
| O | OEM, systems & services provider | Engineering and technology manufacturing |
| G1, G2 | Government and public administration | City planning; economic development |
| I | Insurers | Connected mobility insurance solutions |
| M1 – M3 | Mobility service provider | Public transport; car sharing |
| S | Vulnerable population | Rights and concerns of visually impaired people |

A pool of potential candidates was generated and invited to share their visions in semi-structured interviews. 17 participants, three of which were women, from six European countries, were recruited, with experience in their area of expertise between two and 28 years. Stakeholder categories had been predefined before recruitment as inclusion criteria. No explicit exclusion criteria were defined. No incentives were offered. A letter of information and informed consent were sent a day before the interview.

## 2.2 Semi-structured interviews

Participants were interviewed between July and December 2019 via phone call, and interviews lasted between 30 and 90 mins. Following Patton (2014) and Turner (2010), a general interview guide with predetermined questions was constructed by the three paper authors, who also conducted the interviews. A brief introduction and goal statement led the exploration of stakeholder points of view on autonomous vehicles and vision scenarios for CAV integration; participants were invited to introduce themselves, their position and their experience with autonomous vehicles, then the questionnaire guideline (outlined in Table 2) was employed; the questions were asked almost verbatim and supplied with follow-up questions in case the participants struggled to answer or were unspecific and required clarification.

**Table 2.** Summary of the interview guideline.

| Structure | Questions |
| --- | --- |
| Vision and worst-case scenarios | From your point of view, what is or what are the visions for either connected or autonomous vehicles or both?<br>• What? For whom? Where? When? What are business models / regulations / products?<br>• What would be the positive consequences?<br>• What might be possible negative consequences?<br>• Social? Environmental? Economic?<br>From your point of view, what must NOT happen when it comes to CAVs? |
| Users | Let's talk about the users. Who are the users? What are they doing with the solution?<br>• How would you tell a user accepted the solution?<br>• How would they behave/think/feel?<br>For these behaviors, where do you see acceptance problems?<br>• What do you base this knowledge on?<br>• Do you think users know enough or think they know enough?<br>• Do you think users have the time and money?<br>• Do you think others will allow users to do it? (their parents, children, spouses)<br>• Do you think users are motivated to do it? |

| Others | Let's talk about the acceptance of the solution(s) within your organisation and the larger context. In other words, what might prevent it from becoming reality? |
|---|---|
| | • Does the context - legal, political, economic - allow the solution to be introduced? |
| | • Are there key players for or against the introduction? |
| | Does your organisation have the know-how to aid in CAV introduction? |
| | Do you see privacy and security issues? |

The interview focused on visions solutions and benefits regarding CAVs, as well as worst-case scenarios and risks from the perspective of the participant. Another main target was the prediction of user barriers and motivators upon introduction of CAVs into the mobility eco system; for this section, the participants' vision scenario was utilized as the accepted introduced CAV solution. This was also employed for the discussion on other barriers.

### 2.3    Data recording and analysis

Interviews were annotated into a preformatted guideline sheet by the interviewers and audio recorded with permission of all participants. The audio recordings were then analyzed by two researchers and interviewer notes were supplemented and updated based on the audio recordings. Where discrepancies in interpretation occurred, the audio recording was chosen as the more objective source, and interviewer and transcriber discussed the issue until a consensus was reached. For data analysis, RQDA (Huang, 2014) was employed; data was read into the software and analyzed using thematic analysis (Braun & Clarke, 2006). Based on relevant literature on CAVs, main higher order themes were identified and grouped in a deductive manner. Codes, short phrases that provide meaning in the theoretical context, were then constructed from the themes; an overview can be seen in Figure 1. Additional patterns were inductively deduced from the data and related to previous literature (Patton, 2014). Two researchers discussed the codes for consistency until consensus was reached. Checks were completed using a plenary discussion with a majority of the interview members. The main thematic structure as well as results were presented and validated, while attending experts who had not themselves participated in interviews provided additional validation of content.

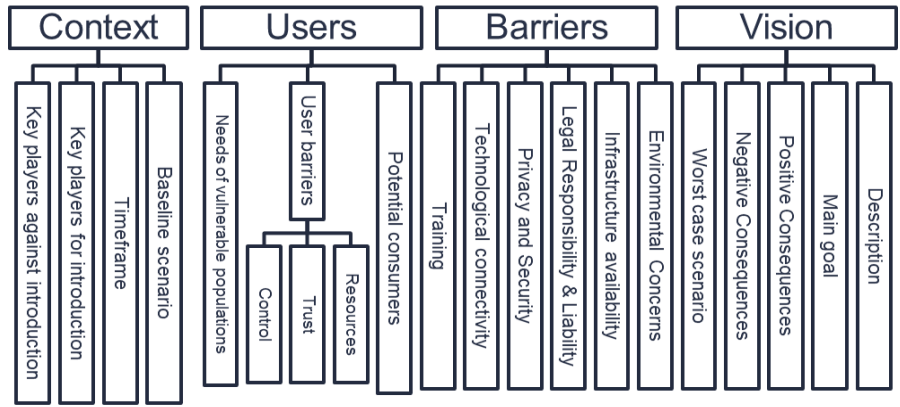

**Fig. 1.** Codes generated during qualitative data analysis

## 3 Results

### 3.1 Visions and worst cases

For the vision scenarios, two major visions emerged: the first, and more commonly mentioned vision, was that CAVs would be sensible in the form of mobility as a service. More specifically, participants named "shuttles for short journeys in demarcated areas such as airports with their own market that would replace today's vehicles" (O), and "shared transport, which is sustainable, as more people will move to the cities and because human driving is highly inefficient" (M1). Vehicles should "not be a propriety item, but ubiquitous in use, where cars are just part of a whole and holistic mobility solution" (I, also C1). It might in this way complement public transport, as it "will not be replaced completely, rather there will be a mix of AVs and public and densely packed autonomous transport" (M1); "the fleet operator will take care of it more than there will be private driving, and you can use it when you need it only" (A1).

The second vision was that CAVs would find introduction within the traditional confines, as privately-owned cars (A1, A3, M1, M2), "with all the issues going along with it, such as climate issues, urban sprawl issues and traffic jams" (C2). Here, multiple participants mentioned that it would first be integrated for "specific tracks on highways, where only some functions will be automated, and where you drive normally and automated only in certain conditions" (A1).

Aside from the two major solutions, some minor other solutions were discussed, such as flying shuttles "with a coordinated takeoff, hybrid electrical, at some point without dedicated pilot" (A2), SMEVs (emergency vehicles) "as connected vehicles that interact with traffic light system, so ambulances or fire brigades have green lights their entire way" (C3), and automated trucks and truck platooning (O). For these solutions, ownership would have to be defined to be either public or corporate.

The worst-case scenario perspective revealed two major themes. One, many stakeholders agreed that in the instance of privately-owned car solution, CAVs "could prevent changes toward what really matters, like active mobility, vehicle sharing, and less convenience" (G1) and "if automation can help, that would be great, but if automation is just another way of giving priority to private car/motorized transport that would be the worst case" (A1). Secondly, "control through external sources" (A3), in the sense that CAVs could result in more external limitations rather than providing more freedom, was discussed - such as through traffic jams, employer and/or government control while in the vehicle, hackers (C4, A3, C2), and "cities built around autonomous vehicles whose routes and parking spaces define how they are built" (A1).

### 3.2 Positive and negative consequences

Six main areas of consequences, labelled (1) to (6) below, were frequently discussed, ranging from very proximal consequences such as comfort to very distal ones, such as ecological sustainability (see Figure 2). They evaluation even of the most distal consequences were marked by ambivalence.

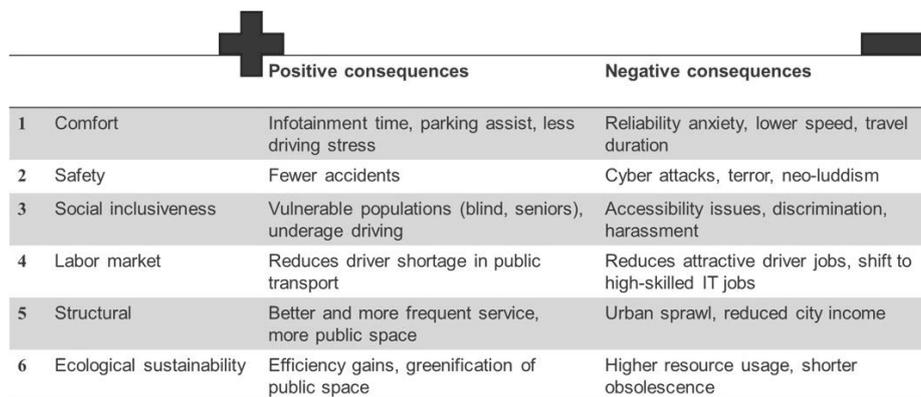

| | | Positive consequences | Negative consequences |
|---|---|---|---|
| 1 | Comfort | Infotainment time, parking assist, less driving stress | Reliability anxiety, lower speed, travel duration |
| 2 | Safety | Fewer accidents | Cyber attacks, terror, neo-luddism |
| 3 | Social inclusiveness | Vulnerable populations (blind, seniors), underage driving | Accessibility issues, discrimination, harassment |
| 4 | Labor market | Reduces driver shortage in public transport | Reduces attractive driver jobs, shift to high-skilled IT jobs |
| 5 | Structural | Better and more frequent service, more public space | Urban sprawl, reduced city income |
| 6 | Ecological sustainability | Efficiency gains, greenification of public space | Higher resource usage, shorter obsolescence |

**Fig. 2.** Illustration of the anticipated positive and negative consequences of the introduction of CAVs; order from proximal aspects such as comfort, to distal ones such as ecological sustainability.

**(1) Comfort**. Most experts expected CAVs to provide "improved activity usage of car time, e.g. working, being entertained, chatting" (A2), the car as infotainment (A2, A3, C2, M1), and "more comfortable smooth rides" with less stress, as "drivers are the weakest part of the driving, due to bad breaking, bad acceleration, and not looking into the future when driving" (M2). Increased comfort would also be provided in the case of SMEVs, as they could "get faster access to patients and to the hospitals, which would lead to lower stress levels for drivers and reduce braking with an eye towards patients" (C3). Searches for parking spaces would also be reduced (G1, C3). Also mentioned was the "choice of various types of vehicles on-call" (C5). Counteracting this increase in comfort, many expected anxieties regarding the proper functioning (C1, S) for example "risk of losing connection when needed" (C1), while some claimed that "experience has to overcome anxiety, experience has to buy acceptance" (A3). Longer travel durations were also mentioned as an issue, "due to speed limits, see the EU regulation 2021, where it says that if you accelerate faster than the speed limit, the car will automatically disengage the accelerator" (C5) and more congestion (O, C1, C5).

**(2) Safety.** Many participants believed that CAVs could improve safety, and that it would for example "cut down on the number of deaths" (C5), as "it will be much safer than human driven vehicles" (M2), among other causes because "alcohol related accidents will be reduced" (A4), and because "people are less likely to break the rules [due to the surveillance]" (C5). An additional safety benefit would be that for example children would get out of the vehicle where they are supposed to get off" (C2). On the other hand, "conflicts and acts of terrorisms are conceivable" (A1), and multiple participants admitted that it would be difficult to lower the risk of hacker attacks (A1, A3, C6, G2) or prevent damage to the system by protesting citizens (I, C4).

**(3) Social inclusiveness.** CAVs were mentioned to be a way to "make people mobile again" (A3), in particular as "the elderly sooner or later cannot drive themselves anymore, and CAVs may help" (C6); the potential for people with visual impairments were also discussed, but creating a proper coverage for all liabilities and possible negative events was considered a difficult topic, as "discrimination against blind people might occur if blind people have a higher incidence rate of accidents because they cannot respond as well to emergencies" (I). Furthermore, CAVs might be worrisome in the context of public transport, as in small spaces, "sexual harassment would be worse, unless everything is recorded (which would only help after the fact), or unless there is a permanent connection with the camera" (M2). This could also lead to discriminatory usage.

**(4) Labor market.** CAVs were discussed as the "solution to the increasing problem of getting good drivers for busses; they can save on costs, and [given recording exists], drivers cannot be robbed" (C4). However, this would have major economic impact (A3, G2), as CAVs overall would lead to "logistics & business drivers no longer needed" (A3) and many people would lose their job, with need for "fewer engineers, more computer scientists" (C2).

**(5) Structural consequences.** CAVs may provide "24/7 mobility, especially in rural areas – and offer for the same money a more comprehensive mobility service both in terms of quantity as well as quality" (C4). On the downside, "this would almost certainly lead to urban sprawl: people could now live in the suburbs because it is cheaper and start working while driving" (C2), and the consequence might be "many empty runs, and even more cars on the road" (G1). Additionally, "parking spaces would be no longer needed" (A3), which could be used for greening projects and allow more space for residents, but would increase CAV driving kms and lower city income from parking fees and fines (G1, C4).

**(6) Environmental consequences.** One might expect that CAVs would lead to "less emissions and less energy consumption, due to the use of more efficient routes, lower congestion and better traffic flow due to high levels of connectivity between vehicles or centralized command" (A3, G1). However, "the increase in personal convenience and potential lower costs is a danger, as it would shift people to use cars instead of buses (M1), i.e. "subtract from public transport" (O). Furthermore "the increase in infrastructure needed" will come at a large cost in electricity (C1) and "obsolescence of vehicles will be quicker due to empty km and continuous driving without parking" (C1, also G1). Additionally, "large amounts of data have to be handled and stored. This requires brutal server capacities, and servers that consume energy. Servers already make up a large share of [global] energy consumption." (C3)

### 3.3 Passengers and their barriers

The user demographic of CAVs was differentiated by scenario; for privately owned vehicles with full automation, participants unanimously agreed that it would be younger people/the younger generations, joined by urban business travelers, and probably at first more male, wealthier, and more educated. Barriers to adoption here would differ somewhat from CAVs adopted for public transport, where "it won't make a difference in terms of demographics" (A1). Three major usage barriers were identified: capability/knowledge, opportunity and motivation, with vulnerable population, people of lower socio-economic status and current car owners respectively being the main exponents of each of the barriers.

**Capability**. Lack of knowledge was discussed as a main barrier to adoption for "older people, and people who are not good at technology and don't want to learn how to use it" (C6). Additionally, for blind people, "confidence is a major concern - if not enough information is available, no backup system in place, blind people will be hesitant to use it" (S). Consumer "confusion due to how manufacturers market AVs (advertisements)" (C5) so that "the "man on the street" has no really good knowledge" (I) was another concern.

**Opportunity.** This barrier was mainly identified for people with lower socio-economic status. "Money is an important determinant; e.g. some people cannot afford a taxi – so if CAVs are also expensive, their problem is not solved" (C6). It is also possible that "CAVs distribution will start on an aggressive price plan in the first phase, but that price will rise with services and time" (G2). Additionally, "connectivity will bring some challenges (e.g. different software)" (C5), which might reduce accessibility for vulnerable populations.

**Motivation**. Here, vehicle owners were discussed as the primary target group. "People who enjoy driving will be hardest to convince to change and will be less willing to accept CAVs" (A), "because driving speed will be regulated" (A4), or because they fear a loss of control (I, A2); a perceived loss of freedom might also lower motivation, because if CAVs are on demand, changes in travel plans are required – questions posed here would be whether "people are willing to share their ride, whether people are willing to wait more than 5-10 minutes for the vehicle, and how far people will be willing to walk" (C1). Cybersecurity, safety perceptions and perceptions of low accessibility might increase this issue further.

### 3.4    Non-passengers and their motivators and barriers

Aside from users, others' interests in adoption or prevention of CAVs were discussed, such as CAVs as an opportunity for data and ride-haling companies (such as Uber/Lyft), and public transport institutions. Here, "human drivers are too expensive, and the biggest overhead of their services can be made cheaper with autonomy" (C1, M2). Companies dealing with information technology, car manufacturers and ministries of economy also stand to gain, especially if CAVs are integrated as personal commodity (A1, G2, C1).

CAVs might not be perceived as so beneficial from the perspective of road co-user associations such as "cyclists and pedestrians, who may criticize the fact that the automotive industry is being further promoted" (C2), and those who advocate the health benefits of these modes of transport. Trade unions (such as bus operations) and "family businesses to medium-sized businesses, that are not yet prepared (as was the case with the e-bus) that will eventually be completely replaced by autonomous vehicles" (C4) were also mentioned.

A major obstacle is seen by multiple participants in politics; on one hand "the sluggishness of regulators is a problem for developing a good system" (C5), especially since "lawmakers in governments could be pushing back due to fear that congestion would get worse" (C1). Secondly, "many city councils are populated by older, wealthier male members, for whom driving cars is a status symbol and deeply ingrained habit, and for whose constituents a focus on cars is emotional, as cars are seen as economic driver and support to prosperity" (G1). Finally, "high cost in the beginning will put off municipalities" (M2), including costs from infrastructure and to the economy in terms of job losses.

## 4    Discussion

At present, connected and autonomous vehicles are being introduced on the streets around the globe (Innamaa, 2019; Lee, 2020). They come in diverse forms. Small shuttle buses extend rail services, ambulances communicate with traffic lights, autonomous vans offer ride-haling services, trucks platoon autonomously on highways, luxury sedans cruise the city streets and flying copters assist their non-pilot users. Scientifically, it is becoming clear that the social, ecological and economic consequences are going to be as diverse as their forms of introduction. Integrated into public transport and mobility-as-a-service, CAVs promise increased energy efficiency, social inclusiveness and livelihood in inner cities. Privately owned, as a means of individual transportation, CAVs might come with more km per vehicle, more vehicles on the roads, more energy usage and less social inclusion.

We analyzed 17 interviews with representatives of stakeholder groups for the introduction of CAVs, including representatives of public administrations, automotive original equipment manufacturers, insurance companies, public transportation service providers, mobility experts and politicians. Their vision and worst-case scenarios reflect the scientific debate. While CAVs that support active and shared mobility were perceived as a major opportunity for sustainable progress, dystopian future visions of CAVs preventing necessary changes away from passive use, and an ownership dominated mobility were at least as prominent.

In line with previous literature on CAVs, our results proved not only heterogeneous, but ambivalent in nature, with positive and negative aspects for virtually every aspect of the integration of CAVs in the mobility eco-system. More specifically, ambivalence was present for each of the six most frequently names categories of consequences. Consequences for comfort, safety, social inclusiveness, the labor market, structural changes and ecological sustainability were all either desirable or undesirable depending on the form of CAV introduction.

Some additional specific aspects deserve pointing out. Other co-road users, such as pedestrians and cyclists, were seen as particularly vulnerable in an early phase of CAV introduction. The needs of the blind and partially sighted, especially with respect to the design of digital interfaces, also seemed underrepresented in current development, in particular considering the magnitude of the positive impact for these groups. Finally, the unprecedented loss of personal freedom and privacy has not received sufficient attention if one considers the scope of information released by CAV usage; some subjects, such the access employers, insurers and marketers might gain to individual movement patterns have been mentioned; other less obvious allowances might yet be revealed. Especially given the heterogeneous consequences, the sluggishness and reactivity of regulators was rightly observed with prominent concern.

The present research has scientific and political implications. When scientifically studying the acceptance of CAVs, care should be taken to clearly specify what form of CAVs is of interest. If determining the form of CAV is left to naïve respondents, results might represent an uninterpretable mixture of ideas about CAVs. Further, given that acceptance will in part reflect the evaluation of consequences, any acceptance measure will critically depend on the information available to respondents. Experimental paradigms seem in order, to allow better study of the contribution of different consequences to acceptance. Political debates should not only focus on whether or not CAVs should be introduced; more importantly, the form of introduction and its implications for a sustainable mobility future need to be at the forefront of the discourse. Since this question touches a multitude of actors, inclusive stakeholder dialogues seem timely.